\documentclass[a4]{JHEP3}
\usepackage{amsmath}

\newcommand{\di}{\text{d}}

\newcommand{\cl}[1]{#1}
\newcommand{\al}[1]{\bar{#1}}

\newcommand{\ci}[1]{{#1}}
\newcommand{\ai}[1]{{\dot{#1}}}

\newcommand{\cd}{\cl{\text{d}}_{\ci{+}}}
\newcommand{\ad}{\al{\text{d}}_{\ai{+}}}

\newcommand{\ddd}[2]{\partial_{\ci{#1}\ai{#2}}}
\newcommand{\dpp}{\ddd{+}{+}}
\newcommand{\dpm}{\ddd{+}{-}}
\newcommand{\dmp}{\ddd{-}{+}}
\newcommand{\dmm}{\ddd{-}{-}}

\newcommand{\dbox}{\Box}

\newcommand{\vdd}[3]{{#1}_{\ci{#2}\ai{#3}}}

\newcommand{\vuu}[3]{{#1}^{\ci{#2}\ai{#3}}}

\newcommand{\caud}[3]{\cl{#1}^{\ci{#2}}{}_{\ci{#3}}}
\newcommand{\cadu}[3]{\cl{#1}_{\ci{#2}}{}^{\ci{#3}}}
\newcommand{\cadd}[3]{\cl{#1}_{\ci{#2}\ci{#3}}}

\newcommand{\aaud}[3]{\al{#1}^{\ai{#2}}{}_{\ai{#3}}}
\newcommand{\aadu}[3]{\al{#1}_{\ai{#2}}{}^{\ai{#3}}}
\newcommand{\aadd}[3]{\al{#1}_{\ai{#2}\ai{#3}}}

\newcommand{\fsn}[1]{#1\kern-0.50em/}

\newcommand{\ppn}[1]{#1}
\newcommand{\ppu}[1]{{#1}_{\ci{-}}}
\newcommand{\ppd}[1]{{#1}_{\ai{-}}}
\newcommand{\ppx}[1]{{#1}_{\ci{-}\ai{-}}}
\newcommand{\ppcn}[1]{\cl{#1}}
\newcommand{\ppcu}[1]{\cl{#1}_{\ci{-}}}

\newcommand{\pptn}[1]{\tilde{#1}}
\newcommand{\pptu}[1]{{\tilde{#1}}_{\ci{-}}}
\newcommand{\pptd}[1]{{\tilde{#1}}_{\ai{-}}}
\newcommand{\pptx}[1]{{\tilde{#1}}_{\ci{-}\ai{-}}}
\newcommand{\pptcn}[1]{\tilde{\cl{#1}}}
\newcommand{\pptcu}[1]{\tilde{\cl{#1}}_{\ci{-}}}
\newcommand{\pptan}[1]{\tilde{\al{#1}}}
\newcommand{\pptad}[1]{\tilde{\al{#1}}_{\ai{-}}}

\newcommand{\ccdi}[1]{\cl{\nabla}_{\ci{#1}}}
\newcommand{\cadi}[1]{\al{\nabla}_{\ai{#1}}}
\newcommand{\ccd}{\ccdi{+}}
\newcommand{\cad}{\cadi{+}}
\newcommand{\cdd}[2]{\nabla_{\ci{#1}\ai{#2}}}
\newcommand{\cdu}[2]{\nabla^{\ci{#1}\ai{#2}}}

\newcommand{\bphi}{\boldsymbol{\phi}}
\newcommand{\bPhi}{\boldsymbol{\Phi}}

\newcommand{\clW}{\boldsymbol{W}}
\newcommand{\clF}{\boldsymbol{F}}
\newcommand{\clD}{\boldsymbol{D}}
\newcommand{\clw}{\boldsymbol{w}}
\newcommand{\clf}{\boldsymbol{f}}
\newcommand{\cld}{\boldsymbol{d}}

\title{Gauge theory and SIM(2) superspace}

\author{Ji\v{r}\'{\i} Voh\'anka\\
Masaryk University, Institute for Theoretical Physics\\
Kotl\'a\v{r}sk\'a 2, 611 37 Brno, Czech Republic\\
Email: \email{vohanka@physics.muni.cz}
}

\abstract{
In this paper the $SIM(2)$ superspace formulation of the supersymmetric Yang-Mills gauge theory minimally coupled to chiral superfields is discussed. The super-Poincare invariant supersymmetric Yang-Mills theory is rewritten to $SIM(2)$ superspace formalism and the effects of $SIM(2)$ invariant but Lorentz breaking terms are discussed. Two approaches are investigated. The first is based on the gauge chiral representation of the supersymmetric gauge theory and the second is based on the covariant representation of the supersymmetric gauge theory.
}


\preprint{}

\begin{document}

 
\section{Introduction}

Cohen and Glashow noticed \cite{cg}, \cite{cg1} that many physical phenomena, like for example the length contraction and time dilatation, are left unchanged if we do not assume the invariance of the physics with respect to the full Lorentz group but only to its $SIM(2)$ subgroup. This opens new possibilities in the particle phenomenology, in particular concerning neutrino masses.

The implications of such a theory for neutrino masses were discussed in \cite{dm}. The modifications of the electromagnetic theory were discussed in \cite{scs}.

The supersymmetric theory based on the $SIM(2)$ subgroup was considered in \cite{cf} and its superspace formulation was developed in \cite{uam}. 
Feynman rules in $SIM(2)$ superspace formalism were presented in \cite{puv}. Wess-Zumino model with Lorentz breaking mass term was used as an example on which the one loop calculation and renormalization was demonstrated.

If we wish to consider neutrions with mass added by a $SIM(2)$ invariant but Lorentz breaking term, then the whole standard model, including the gauge sector, has to be treated as a $SIM(2)$ symmetric theory. Thus it is important to investigate the implications of $SIM(2)$ symmetry for gauge theories. In the non-supersymmetric case this was done for example in \cite{scs} and \cite{dm}.

In the supersymmetric case there are several articles, for example \cite{cf}, \cite{uam}, \cite{puv}, where the detailed treatment of a theory with chiral multiplet is provided. Although the aspects of the gauge theory are discussed on several ocasions in these articles and the $SIM(2)$ modifications of it are provided, the detailed treatment of gauge theory, especially in a superspace formulation is still missing. This article tries to fill this gap.

The paper is organised as follows. 
In sections \ref{sec:sim2} and \ref{sec:sim2susy} the notion of $SIM(2)$ (super)field is introduced and it is shown how Lorentz (super)field can be decomposed into $SIM(2)$ (super)fields. In order to gain familiarity with $SIM(2)$ (super)field formalism, the equations of motion for massless spin $\tfrac{1}{2}$ and spin $2$ fields are rewritten using $SIM(2)$ fields and discussed in detail. 
The sections \ref{sec:chiab}, \ref{sec:chiym}, \ref{sec:chiel} are devoted to the gauge chiral representation of supersymmetric gauge theory. In section \ref{sec:chiab} the simple case of abelian gauge theory is discussed, it is shown that gauge freedom can be completely fixed in $SIM(2)$ invariant way. The results are then generalised to non-abelian case in section \ref{sec:chiym} and in section \ref{sec:chiel} it is shown that some of the $SIM(2)$ superfields are auxiliary and can be eliminated from the action.
Section \ref{sec:cov} is devoted to the covariant representation of supersymmetric gauge theory. At the end of this section it is shown how the covariant representation is related to the gauge chiral representation presented in previous sections.
In section \ref{sec:break} it is discussed how are the results presented in previous sections are affected if we add a $SIM(2)$ invariant but Lorentz breaking mass term.


\section{$SIM(2)$ group and properties of $SIM(2)$ fields}
\label{sec:sim2}

$SIM(2)$ is a subgroup of the Lorentz group which preserves a chosen null vector $n$ up to rescalings. We will assume that this null vector is chosen such that its coordinates are $\vuu{n}{+}{+}=1$, $\vuu{n}{+}{-}=\vuu{n}{-}{+}=\vuu{n}{-}{-}=0$.

The $SIM(2)$ group is four dimensional and solvable. As a basis of its Lie algebra we can choose four generators of symmetry $\cadd{J}{+}{+}$, $\cadd{J}{+}{-}$, $\aadd{J}{+}{+}$, $\aadd{J}{+}{-}$. Because it is solvable, we know from the theory of group representations, that all irreducible representations are one dimensional, but not all of its representations are fully reducible.

Our main purpose will be to modify a Lorentz invariant theory by adding small $SIM(2)$ invariant but not Lorentz invariant perturbations. For this reason we are not interested in the general theory of representations of the $SIM(2)$ group. It will be enough for us to look at how the representations of the Lorentz group behave when we reduce the symmetry to only the $SIM(2)$ subgroup.

We start by looking at the behaviour of left and right handed Weyl spinors (i.e. representations $(\tfrac{1}{2},0)$ and $(0,\tfrac{1}{2})$). We can decompose the spinor space by the method described in \cite{uam}. We introduce another null vector $\tilde{n}$ satisfying $n\cdot\tilde{n}=1$, whose components we choose to be $\vdd{\tilde{n}}{+}{+}=1$, $\vdd{\tilde{n}}{+}{-}=\vdd{\tilde{n}}{-}{+}=\vdd{\tilde{n}}{-}{-}=0$. Then the spinor space can be decomposed by the projectors $\frac{\fsn{n}\fsn{\tilde{n}}}{2}$ and $\frac{\fsn{\tilde{n}}\fsn{n}}{2}$. The projector $\frac{\fsn{n}\fsn{\tilde{n}}}{2}$ projects the left(right) Weyl spinors on the one-dimensional $SIM(2)$ invariant subspace, while the projector $\frac{\fsn{\tilde{n}}\fsn{n}}{2}$ projects on its one-dimensional complement, which is not uniquely determined, because there is a freedom in choosing $\tilde{n}$.
\footnote{
In our particular choice of $n$ and $\tilde{n}$ the left and right Weyl spinors are decomposed as
\begin{align}
 \cl{\psi}& = \left(\begin{matrix}\cl{\psi}_\ci{+}\\\cl{\psi}_\ci{-}\end{matrix}\right) 
  = \left(\begin{matrix}0\\ \cl{\psi}_\ci{-}\end{matrix}\right) + \left(\begin{matrix}\cl{\psi}_\ci{+}\\ 0\end{matrix}\right), &
 \al{\psi} &= \left(\begin{matrix}\al{\psi}_\ai{+}\\ \al{\psi}_\ai{-}\end{matrix}\right)
  = \left(\begin{matrix}0\\ \al{\psi}_\ai{-}\end{matrix}\right) + \left(\begin{matrix}\al{\psi}_\ai{+}\\ 0\end{matrix}\right),
\end{align}
where the first term on the right hand side belongs to the invariant subspace, while the second belongs to its complement.
}
The action of infinitesimal $SIM(2)$ rotations on the Weyl spinors is
\begin{align}
\label{eq:sim2:sim2_rotations_matrix}
 \begin{pmatrix} \delta\cl{\psi}_\ci{+} \\ \delta\cl{\psi}_\ci{-} \end{pmatrix} &=
  \begin{pmatrix} i\cadd{\epsilon}{+}{-} & 0 \\ i\cadd{\epsilon}{-}{-} & -i\cadd{\epsilon}{+}{-}\end{pmatrix}
  \begin{pmatrix} \cl{\psi}_\ci{+} \\ \cl{\psi}_\ci{-} \end{pmatrix},
&
 \begin{pmatrix} \delta\al{\psi}_\ai{+} \\ \delta\al{\psi}_\ai{-} \end{pmatrix} &=
  \begin{pmatrix} i\aadd{\epsilon}{+}{-} & 0 \\ i\aadd{\epsilon}{-}{-} & -i\aadd{\epsilon}{+}{-}\end{pmatrix}
  \begin{pmatrix} \al{\psi}_\ai{+} \\ \al{\psi}_\ai{-} \end{pmatrix},
\end{align}
where $\cadd{\epsilon}{+}{-}$, $\cadd{\epsilon}{-}{-}$, $\aadd{\epsilon}{+}{-}$, $\aadd{\epsilon}{-}{-}$ are some infinitesimal parameters. 

If we want to know how other finite dimensional representations of the Lorentz group behave when we reduce the symmetry to the $SIM(2)$ subgroup, we use the fact that any such representation can be expressed in terms of tensor products of left and right handed Weyl representations, whose behaviour have already been discussed. 

Now we look at how the fields, i.e. functions over configuration space carrying representation of the Lorentz group, behave when we reduce the symmetry from the Lorentz group to the $SIM(2)$ subgroup. The simplest case is a scalar field, which transforms as $\phi'(x')=\phi(x)$. In the infinitesimal form this reads as
\begin{equation}
\label{eq:sim2:scalar_infinit_transf}
 \delta\phi(x) = -\delta\vuu{x}{\alpha}{\alpha}\ddd{\alpha}{\alpha}\phi(x),
\end{equation}
where the infinitesimal transformations of space time coordinates are
\begin{equation}
\label{eq:sim2:coord_infinit_transf}
 \delta\vuu{x}{\alpha}{\alpha} = 
  -\caud{\epsilon}{\alpha}{\beta}\vuu{x}{\beta}{\alpha} - \aaud{\epsilon}{\alpha}{\beta}\vuu{x}{\alpha}{\beta}.
\end{equation}

When the symmetry is reduced to the $SIM(2)$ subgroup, the rule \eqref{eq:sim2:scalar_infinit_transf} remains valid but the infinitesimal transformations of space time coordinates $\delta\vuu{x}{\alpha}{\alpha}$ are less general because $\caud{\epsilon}{-}{+}$ and $\aaud{\epsilon}{-}{+}$ are set to zero in \eqref{eq:sim2:coord_infinit_transf}.

The case of spinor fields, which transforms under the infinitesimal Lorentz rotations as
\begin{align}
 \delta\cl{\psi}_\ci{\alpha}(x) &= -\delta\vuu{x}{\alpha}{\alpha}\ddd{\alpha}{\alpha}\cl{\psi}_\ci{\alpha}(x) + \cadu{\epsilon}{\alpha}{\beta}\cl{\psi}_\ci{\beta}(x), \nonumber\\
 \delta\al{\psi}_\ai{\alpha}(x) &= -\delta\vuu{x}{\alpha}{\alpha}\ddd{\alpha}{\alpha}\al{\psi}_\ai{\alpha}(x) + \aadu{\epsilon}{\alpha}{\beta}\al{\psi}_\ai{\beta}(x),
\end{align}
is far more interesting. When the symmetry is reduced to the $SIM(2)$ subgroup, then we use the projectors $\frac{\fsn{n}\fsn{\tilde{n}}}{2}$ and $\frac{\fsn{\tilde{n}}\fsn{n}}{2}$ to split the fields in the same way as we did in the case of the left and right Weyl spinors. The infinitesimal $SIM(2)$ transformations now read as
\begin{align}
\label{eq:sim2:spinor_sim2rot_nonredefined}
 \delta\cl{\psi}_\ci{+}(x) &= -\delta\vuu{x}{\alpha}{\alpha}\ddd{\alpha}{\alpha}\cl{\psi}_\ci{+}(x) + i\cadd{\epsilon}{+}{-}\cl{\psi}_\ci{+}(x), \nonumber\\
 \delta\cl{\psi}_\ci{-}(x) &= -\delta\vuu{x}{\alpha}{\alpha}\ddd{\alpha}{\alpha}\cl{\psi}_\ci{-}(x) - i\cadd{\epsilon}{+}{-}\cl{\psi}_\ci{-}(x) 
  + i\cadd{\epsilon}{-}{-}\cl{\psi}_\ci{+}(x), \nonumber\\
 \delta\al{\psi}_\ai{+}(x) &= -\delta\vuu{x}{\alpha}{\alpha}\ddd{\alpha}{\alpha}\al{\psi}_\ai{+}(x) + i\aadd{\epsilon}{+}{-}\al{\psi}_\ai{+}(x), \nonumber\\
 \delta\al{\psi}_\ai{-}(x) &= -\delta\vuu{x}{\alpha}{\alpha}\ddd{\alpha}{\alpha}\al{\psi}_\ai{-}(x) - i\aadd{\epsilon}{+}{-}\al{\psi}_\ai{-}(x) 
  + i\aadd{\epsilon}{-}{-}\al{\psi}_\ai{+}(x).
\end{align}
While the transformation of the projection $\cl{\psi}_\ci{+}(x)$ is expressed in terms of itself and does not depend on the projection $\cl{\psi}_\ci{-}(x)$, the transformation of the projection $\cl{\psi}_\ci{-}(x)$ depends on both $\cl{\psi}_\ci{+}(x)$ and $\cl{\psi}_\ci{-}(x)$. We cannot separate the projections $\cl{\psi}_\ci{+}(x)$ and $\cl{\psi}_\ci{-}(x)$ from each other because they are mixed by $SIM(2)$ transformations.
This means that in a $SIM(2)$ symmetric theory, the field $\cl{\psi}_\ci{+}(x)$ may appear without $\cl{\psi}_\ci{-}(x)$ being present, but $\cl{\psi}_\ci{-}(x)$ has to appear in multiplet with $\cl{\psi}_\ci{+}(x)$.
The same is true also for $\al{\psi}_\ai{+}(x)$ and $\al{\psi}_\ai{-}(x)$. 
However we can remedy this by defining modified projections $\tilde{\cl{\psi}}_\ci{+}(x)$, $\tilde{\cl{\psi}}_\ci{-}(x)$ and $\tilde{\al{\psi}}_\ai{+}(x)$, $\tilde{\al{\psi}}_\ai{-}(x)$, which have the property that $SIM(2)$ transformations do not mix them among each other. They are defined as
\begin{align}
\label{eq:sim2:spinor_redefinition}
 \tilde{\cl{\psi}}_\ci{+}(x) &= \cl{\psi}_\ci{+}(x), &
 \tilde{\cl{\psi}}_\ci{-}(x) &= \cl{\psi}_\ci{-}(x) - \frac{\dmp}{\dpp}\cl{\psi}_\ci{+}(x), \nonumber\\
 \tilde{\al{\psi}}_\ai{+}(x) &= \al{\psi}_\ai{+}(x), &
 \tilde{\al{\psi}}_\ai{-}(x) &= \al{\psi}_\ai{-}(x) - \frac{\dpm}{\dpp}\al{\psi}_\ai{+}(x),
\end{align}
and their infinitesimal transformations are
\begin{align}
\label{eq:sim2:spinor_sim2rot_redefined}
 \delta\tilde{\cl{\psi}}_\ci{+}(x) &= -\delta\vuu{x}{\alpha}{\alpha}\ddd{\alpha}{\alpha}\tilde{\cl{\psi}}_\ci{+}(x) + i\cadd{\epsilon}{+}{-}\tilde{\cl{\psi}}_\ci{+}(x), \nonumber\\
 \delta\tilde{\cl{\psi}}_\ci{-}(x) &= -\delta\vuu{x}{\alpha}{\alpha}\ddd{\alpha}{\alpha}\tilde{\cl{\psi}}_\ci{-}(x) - i\cadd{\epsilon}{+}{-}\tilde{\cl{\psi}}_\ci{-}(x), \nonumber\\
 \delta\tilde{\al{\psi}}_\ai{+}(x) &= -\delta\vuu{x}{\alpha}{\alpha}\ddd{\alpha}{\alpha}\tilde{\al{\psi}}_\ai{+}(x) + i\aadd{\epsilon}{+}{-}\tilde{\al{\psi}}_\ai{+}(x), \nonumber\\
 \delta\tilde{\al{\psi}}_\ai{-}(x) &= -\delta\vuu{x}{\alpha}{\alpha}\ddd{\alpha}{\alpha}\tilde{\al{\psi}}_\ai{-}(x) - i\aadd{\epsilon}{+}{-}\tilde{\al{\psi}}_\ai{-}(x).
\end{align}
Because the $SIM(2)$ transformations do not mix them among each other we can regard each of them as a separate $SIM(2)$ field, each of them may appear separately in a $SIM(2)$ symmetric theory. 
Unlike the case with unmodified projections, $\tilde{\cl{\psi}}_\ci{-}(x)$ does not have to appear in the multiplet with $\tilde{\cl{\psi}}_\ci{+}(x)$.
Note that apart from the change of the $x$ variable, the transformations of $\tilde{\cl{\psi}}_\ci{+}(x)$ and $\tilde{\cl{\psi}}_\ci{-}(x)$ are governed only by one (complex) parameter $\cadd{\epsilon}{+}{-}$. This parameter scales and changes the phase of these $SIM(2)$ fields in such a way that the scale and the phase of $\tilde{\cl{\psi}}_\ci{-}(x)$ is changed in the opposite way as for $\tilde{\cl{\psi}}_\ci{+}(x)$. This gives us a nice interpretation of the subscripts $+$ and $-$.
\footnote{
This fact is useful when we are constructing $SIM(2)$ invariants. For example it is easy to understand why the expression $\int\di^4x\tilde{\cl{\psi}}_\ci{-}(x)\dpp\tilde{\al{\psi}}_\ai{-}(x)$ is $SIM(2)$ invariant. The derivative $\dpp$ transforms as 
$\dpp'=(1+i\cadd{\epsilon}{+}{-}+i\aadd{\epsilon}{+}{-})\dpp$ so it is scaled by both parameters $\cadd{\epsilon}{+}{-}$ and $\aadd{\epsilon}{+}{-}$, while each of the $SIM(2)$ fields $\tilde{\cl{\psi}}_\ci{-}(x)$ and $\tilde{\al{\psi}}_\ai{-}(x)$ is scaled by one of them in the opposite way as $\dpp$. The result is that the expression $\tilde{\cl{\psi}}_\ci{-}(x)\dpp\tilde{\al{\psi}}_\ai{-}(x)$ is not scaled at all. The integral ensures that the expression is invariant with respect to the transformations of the $x$ variable so the whole expression is $SIM(2)$ invariant.
}

The price we have to pay for the nice properties of these $SIM(2)$ fields is the introduction of the nonlocal operator $\tfrac{1}{\dpp}$. This operator has to be linear, has to satisfy the condition $\dpp\tfrac{1}{\dpp}=1$, which defines it as a Green function of $\dpp$ and we will also require that it commutes with all space-time derivatives $\left[\ddd{\alpha}{\alpha},\tfrac{1}{\dpp}\right]=0$.
Let $f(x)$ be some function, then the condition that the derivation $\dpp$ has to commute with $\frac{1}{\dpp}$ gives
\begin{equation}
\label{eq:sim2:nlop_function_condition}
 \left[\frac{1}{\dpp},\dpp\right]f(x) = \left(\frac{1}{\dpp}\dpp - \dpp\frac{1}{\dpp}\right)f(x) = \frac{1}{\dpp}\dpp f(x) - f(x) = 0.
\end{equation}
But this is evidently not true for nonzero functions satisfying $\dpp f(x)=0$. This indicates that we have to work with the space of functions which is somewhat reduced, namely to those satisfying \eqref{eq:sim2:nlop_function_condition}. One way how to define the operator $\frac{1}{\dpp}$ is
\begin{equation}
 \frac{1}{\dpp}f(x) = \int_{-\infty}^{\vuu{x}{+}{+}}\di\vuu{t}{+}{+}f(\vuu{t}{+}{+}).
\end{equation}
In this case the space of functions we are working with has to be reduced to those satisfying $\lim_{\vuu{x}{+}{+}\rightarrow -\infty}f(x)=0$.
One of the most important consequences of the fact that we have to work with the reduced space of functions is that the equation $\dpp f(x)=0$ has only one solution $f(x)=0$.

In order to understand the behaviour of these $SIM(2)$ fields and their relation to the Lorentz fields from which we constructed them we will look at two well known models. First we will look at a massless fermion and then at an abelian gauge field.

A massless fermion is described by a spinor field $\cl{\psi}_\ci{\alpha}(x)$ satisfying the equation of motion
\begin{equation}
 \ddd{\alpha}{\alpha}\cl{\psi}^\ci{\alpha}(x)=0.
\end{equation}
If we rewrite it in terms of $SIM(2)$ fields $\tilde{\cl{\psi}}_\ci{+}(x)$ and  $\tilde{\cl{\psi}}_\ci{-}(x)$ we get a set of equations
\begin{align}
 i\dpp\tilde{\cl{\psi}}_\ci{-}(x) &= 0, &
 i\dpm\tilde{\cl{\psi}}_\ci{-}(x) - \frac{\dbox}{i\dpp}\tilde{\cl{\psi}}_\ci{+}(x) &= 0.
\end{align}
As was mentioned before, the equation $\dpp\tilde{\cl{\psi}}_\ci{-}(x)=0$ implies $\tilde{\cl{\psi}}_\ci{-}(x)=0$ because we are forced to work with the restricted space of functions. Thus the above equations are equivalent to
\begin{align}
\label{eq:sim2:eom_ferminon_redefined}
 \tilde{\cl{\psi}}_\ci{-}(x) &= 0, &
 \dbox\tilde{\cl{\psi}}_\ci{+}(x) &= 0.
\end{align}
We see that all dynamics is carried by the field $\tilde{\cl{\psi}}_\ci{+}(x)$, while the field $\tilde{\cl{\psi}}_\ci{-}(x)$ is auxiliary.

An abelian gauge field $A_{a}(x)$ (or equivalently $\vdd{A}{\alpha}{\alpha}(x)$ in the spinor notation) is a vector field, which is subject to the gauge transformation 
\begin{equation}
 A'_{a}(x) = A_{a}(x) + \partial_{a}g(x),
\end{equation}
where $g(x)$ is an arbitrary scalar function. The equation of motion is
\begin{equation}
\label{eq:sim2:maxwell_lorentz}
 \partial^{a}\left( \partial_{a}A_{b}(x) - \partial_{b}A_{a}(x) \right) = 0.
\end{equation}
We will work in the light-cone gauge $n\cdot A(x)=0$, which breaks Lorentz invariance, but does not break $SIM(2)$ invariance. This condition does not fix the gauge completely, we can still perform the gauge transformations with $n\cdot\partial g(x)=0$. 
If we work on-shell we can use this gauge freedom to set $\partial\cdot A(x)=0$. In order to do that we have to perform a gauge transformation with function $g(x)$ satisfying the set of equations $\dbox g(x)=-\partial\cdot A(x)$, $n\cdot\partial g(x)=0$. This set of equations has solution only if 
$\dbox (n\cdot\partial) g(x)=(n\cdot\partial)\dbox g(x)=-n\cdot\partial(\partial\cdot A(x))=0$, but the validity of this integrability condition is ensured by equation of motion as can be easily verified by multiplying \eqref{eq:sim2:maxwell_lorentz} by $n^b$. Thus we can search for solutions satisfying
\begin{align}
 n\cdot A(x) &= 0, & \partial\cdot A(x) &= 0, & \dbox A(x) &= 0.
\end{align}

Now we look how the light-cone gauge and equations of motion look like if we work with $SIM(2)$ fields
\begin{align}
\label{eq:sim2:maxwell_sim2_fields}
 \vdd{\tilde{A}}{+}{+}(x) &= \vdd{A}{+}{+}(x), \nonumber\\
 \vdd{\tilde{A}}{-}{+}(x) &= \vdd{A}{-}{+}(x) - \frac{\dmp}{\dpp}\vdd{A}{+}{+}(x), \nonumber\\
 \vdd{\tilde{A}}{+}{-}(x) &= \vdd{A}{+}{-}(x) - \frac{\dpm}{\dpp}\vdd{A}{+}{+}(x), \nonumber\\
 \vdd{\tilde{A}}{-}{-}(x) &= \vdd{A}{-}{-}(x) - \frac{\dmp}{\dpp}\vdd{A}{+}{-}(x) - \frac{\dpm}{\dpp}\vdd{A}{-}{+}(x) + \frac{\dmp}{\dpp}\frac{\dpm}{\dpp}\vdd{A}{+}{+}(x),
\end{align}
which are defined in such a way that each of them is closed under the action of $SIM(2)$ group. In the light-cone gauge $\vdd{\tilde{A}}{+}{+}(x)=0$ and we are allowed to make gauge transformations with $\dpp g(x)=0$. However we are forced to work with reduced space of functions where the equation $\dpp g(x)=0$ allows only one solution $g(x)=0$. This means that in this formalism the light-cone gauge completely fixes the gauge freedom. The equations of motion are now
\begin{align}
 \dpp\dpp\vdd{\tilde{A}}{-}{-}(x) & = 0, \nonumber\\
 \dbox\vdd{\tilde{A}}{-}{+}(x) + \dmp\dpp\vdd{\tilde{A}}{-}{-}(x) &= 0, \nonumber\\
 \dbox\vdd{\tilde{A}}{+}{-}(x) + \dpm\dpp\vdd{\tilde{A}}{-}{-}(x) &= 0, \nonumber\\
 \frac{\dmp}{\dpp}\dbox\vdd{\tilde{A}}{+}{-}(x) + \frac{\dpm}{\dpp}\dbox\vdd{\tilde{A}}{-}{+}(x) + \dmp\dpm\vdd{\tilde{A}}{-}{-}(x) &= 0.
\end{align}
Because the equation $\dpp\dpp\vdd{\tilde{A}}{-}{-}(x)=0$ has only one solution $\vdd{\tilde{A}}{-}{-}(x)=0$, we have to search for solutions satisfying
\begin{align}
\label{eq:sim2:eom_maxwell}
 \vdd{\tilde{A}}{+}{+}(x) &= 0, & \vdd{\tilde{A}}{-}{-}(x) &= 0, & \dbox\vdd{\tilde{A}}{-}{+}(x) &= 0 = \dbox\vdd{\tilde{A}}{+}{-}(x).
\end{align}
We see that all dynamics is carried by the complex field $\vdd{\tilde{A}}{-}{+}(x)$, while the field $\vdd{\tilde{A}}{-}{-}(x)$ is auxiliary. Although the calculations were affected by the fact that the space of functions is reduced, we still get the correct number of physical modes for the Maxwell equations of motion.

\section{$SIM(2)$ supergroup and properties of $SIM(2)$ superfields}
\label{sec:sim2susy}

The Lie superalgebra of $SIM(2)$ supersymmetry \cite{cf}, is obtained by reducing the super-Poincare superalgebra. The Lorentz part is reduced to $SIM(2)$ rotations and the supertranslations $\cl{\varepsilon}\cl{Q} + \al{\varepsilon}\al{Q}$ are restricted to those satisfying $\fsn{n}\cl{\varepsilon} = 0 = \fsn{n}\al{\varepsilon}$.

The $SIM(2)$ superspace and its algebra of covariant derivatives \cite{uam} can be obtained from their super-Poincare counterparts. The $SIM(2)$ superspace has all of the space-time coordinates but the set of Grassmann odd coordinates is reduced to the projection $\frac{\fsn{n}\fsn{\tilde{n}}}{2}\cl{\theta}$ and its hermitian conjugate.
The covariant spinor derivative $\cl{D}$ can be split into the piece $\cl{d}=\frac{\fsn{\tilde{n}}\fsn{n}}{2}\cl{D}$ which we keep in the algebra and the remaining piece $\cl{q}=\frac{\fsn{n}\fsn{\tilde{n}}}{2}\cl{D}$.

If we use our choice of vectors $n$, $\tilde{n}$, the generators of the $SIM(2)$ superalgebra consists of rotations $\cadd{J}{+}{+}$, $\cadd{J}{+}{-}$, $\aadd{J}{+}{+}$, $\aadd{J}{+}{-}$, space-time translations $\vdd{P}{\alpha}{\alpha}$ and supertranslations $\cl{Q}_\ci{+}$, $\al{Q}_\ai{+}$. The superspace is parametrised by space-time coordinates $\vuu{x}{\alpha}{\alpha}$ and two Grassmann odd coordinates $\cl{\theta}^\ci{+}$, $\al{\theta}^\ai{+}$. The algebra of covariant derivatives consists of derivatives $\cd=\cl{D}_\ci{+}$, $\ad=\al{D}_\ai{+}$ (the minus components of $\cl{d}=\frac{\fsn{\tilde{n}}\fsn{n}}{2}\cl{D}$ and $\al{d}=\frac{\fsn{\tilde{n}}\fsn{n}}{2}\al{D}$ are equal to zero) which are subject to the relation
\begin{equation}
\label{eq:sim2:d_algebra}
 \{\cd,\ad\} = i\dpp.
\end{equation}
The only nonzero components of $\cl{q}$ and $\al{q}$ are $\cl{q}_\ci{-}=\cl{D}_\ci{-}$ and $\al{q}_\ai{-}=\al{D}_\ai{-}$.

In order to rewrite actions and other expressions containing super-Poincare invariant superfields into $SIM(2)$ formalism we need to replace each super-Poincare superfield by a set of $SIM(2)$ superfields having the same component content. This is done by the method of covariant projections. First define a projection symbol
\begin{equation}
 \rVert \equiv \vert_{\frac{\fsn{\tilde{n}}\fsn{n}}{2}\cl{\theta}=0,\frac{\fsn{\tilde{n}}\fsn{n}}{2}\al{\theta}=0} = \vert_{\cl{\theta}^\ci{-}=0,\al{\theta}^\ai{-}=0},
\end{equation}
which projects the Grassmann odd coordinates which are not part of the $SIM(2)$ superspace to zero.

We replace a scalar complex super-Poincare superfield $F$ with four complex $SIM(2)$ projections $\ppn{f}$, $\ppu{f}$, $\ppd{f}$, $\ppx{f}$ related to the superfield $F$ as
\begin{align}
 \ppn{f} &= F\rVert, \nonumber\\
 \ppu{f} &= \cl{q}_\ci{-}F\rVert, \nonumber\\
 \ppd{f} &= \al{q}_\ai{-}F\rVert, \nonumber\\
 \ppx{f} &= \tfrac{1}{2}\left[\cl{q}_\ci{-},\al{q}_\ai{-}\right]F\rVert.
\end{align}
The super-Poincare superfields will be denoted by uppercase letters, while its $SIM(2)$ projections will be denoted by the same lowercase letter.
When the the projections $\ppn{f}$, $\ppu{f}$, $\ppd{f}$, $\ppx{f}$ are $SIM(2)$ rotated, they are mixed with each other.
This is a consequence of the fact that spinor covariant derivatives are transformed according to \eqref{eq:sim2:spinor_sim2rot_nonredefined} under $SIM(2)$ rotations. The minus components of the covariant derivatives, which are used to define the projections, are mixed with the plus components which results in the mixing of projections.

In the case of a chiral superfield $\cl{\Phi}$, only two projections $\ppn{\cl{\phi}}$, $\ppu{\cl{\phi}}$ are independent, in the case of antichiral superfield $\al{\Phi}$ only the projections $\ppn{\al{\phi}}$, $\ppd{\al{\phi}}$ are independent
\footnote{
The other projections are
\begin{align}
 \ppd{\cl{\phi}} &= 0, & \ppx{\cl{\phi}} &= -\tfrac{i}{2}\dmm\ppn{\cl{\phi}}, &
 \ppu{\al{\phi}} &= 0, & \ppx{\al{\phi}} &= \tfrac{i}{2}\dmm\ppn{\al{\phi}}.
\end{align}
}
, moreover they satisfy the conditions
\begin{align}
\label{eq:sim2:chirality_cond_nonredefined}
 \ad\ppn{\cl{\phi}} &= 0, & \ad\ppu{\cl{\phi}} &= i\dmp\ppn{\cl{\phi}}, \nonumber\\
 \cd\ppn{\al{\phi}} &= 0, & \cd\ppd{\al{\phi}} &= i\dpm\ppn{\al{\phi}}.
\end{align}

\newcommand{\shiftop}{-\left(\delta\vuu{x}{\alpha}{\alpha}\ddd{\alpha}{\alpha} + \delta\cl{\theta}^\ci{+}\cl{\partial}_\ci{+} + \delta\al{\theta}^\ai{+}\al{\partial}_\ai{+}\right)}

It is possible to change the definition of the projections in such a way, that $SIM(2)$ rotations do not mix them among each other, moreover we will see, that in the case of a chiral superfield the conditions \eqref{eq:sim2:chirality_cond_nonredefined} will be simplified. The idea is that we replace the covariant derivatives used in the definition of $SIM(2)$ projections with operators defined according to \eqref{eq:sim2:spinor_redefinition}, i.e. with the operators
\begin{align}
\label{eq:sim2:D_redefined}
 \tilde{\cl{q}}_\ci{-} &= \cl{D}_\ci{-} - \frac{\dmp}{\dpp}\cl{D}_\ci{+}, & 
 \tilde{\al{q}}_\ai{-} &= \al{D}_\ai{-} - \frac{\dpm}{\dpp}\al{D}_\ai{+}.
\end{align}
The new projections are defined as
\begin{align}
\label{eq:sim2:projections_redefined}
 \pptn{f} &= F\rVert, \nonumber\\
 \pptu{f} &= \tilde{\cl{q}}_\ci{-}F\rVert, \nonumber\\
 \pptd{f} &= \tilde{\al{q}}_\ai{-}F\rVert, \nonumber\\
 \pptx{f} &= \tfrac{1}{2}\left[\tilde{\cl{q}}_\ci{-},\tilde{\al{q}}_\ai{-}\right]F\rVert.
\end{align}
They transform by rescaling under $SIM(2)$ group.
The infinitesimal $SIM(2)$ rotations are
\begin{align}
\label{eq:sim2:rotations_redefined}
 \delta\pptn{f} &= \shiftop\pptn{f}, \nonumber\\
 \delta\pptu{f} &= \shiftop\pptu{f} - i\cadd{\epsilon}{+}{-}\pptu{f}, \nonumber\\
 \delta\pptd{f} &= \shiftop\pptd{f} - i\aadd{\epsilon}{+}{-}\pptd{f}, \nonumber\\
 \delta\pptx{f} &= \shiftop\pptx{f} - i(\cadd{\epsilon}{+}{-} + \aadd{\epsilon}{+}{-})\pptx{f}.
\end{align}
The first term on the right side accounts for the shift in the coordinates, the second term results in the scaling.

In the case of a chiral superfield $\cl{\Phi}$ only the projections $\pptcn{\phi}$, $\pptcu{\phi}$ are nonzero, for an  antichiral superfield $\al{\Phi}$ only $\pptan{\phi}$, $\pptad{\phi}$ are nonzero, moreover they satisfy the conditions
\begin{align}
\label{eq:sim2:chiral_conditions}
 \ad\pptcn{\phi} &= 0, & \ad\pptcu{\phi} &= 0, \nonumber\\
 \cd\pptan{\phi} &= 0, & \cd\pptad{\phi} &= 0.
\end{align}
We will call $SIM(2)$ superfields satisfying such conditions $SIM(2)$ chiral and $SIM(2)$ antichiral.
The following holds for the hermitian conjugation
\begin{align}
\label{eq:sim2:chiral_conjugation}
 \overline{(\pptcn{\phi})} &= \pptan{\phi}, & \overline{(\pptcu{\phi})} &= -\pptad{\phi}, & 
 \overline{(\pptan{\phi})} &= \pptcn{\phi}, & \overline{(\pptad{\phi})} &= -\pptcu{\phi}.
\end{align}

In the case of real superfield $V = \bar{V}$ the hermitian conjugation acts as
\begin{align}
\label{eq:sim2:real_scalar_conjugation}
 \overline{(\pptn{v})} &= \pptn{v}, & \overline{(\pptu{v})} &= -\pptd{v}, & \overline{(\pptd{v})} &= -\pptu{v}, & \overline{(\pptx{v})} &= \pptx{v}.
\end{align}

One of the remarkable properties of $SIM(2)$ superfields is that any complex $SIM(2)$ superfield $f$ can be decomposed as a sum of a chiral $SIM(2)$ superfield $\cl{c}$ and an antichiral $SIM(2)$ superfield $\al{a}$
\begin{equation}
\label{eq:sim2:decomposition}
 f = \al{a} + \cl{c}.
\end{equation}
The chiral and antichiral $SIM(2)$ superfields appearing in the decomposition can be calculated as
\begin{align}
 \cl{c} &= \frac{\ad\cd}{i\dpp}f, & \al{a} &= \frac{\cd\ad}{i\dpp}f,
\end{align}
the identity \eqref{eq:sim2:d_algebra} then leads to \eqref{eq:sim2:decomposition}.


\section{Abelian case}
\label{sec:chiab}

The aim of this section is to show how does the super-Poincare symmetric abelian gauge theory looks like if we rewrite it in the $SIM(2)$ formalism. We have decided to treat the abelian case separately from the non-abelian case because its simplicity allows us to perform calculations which would be difficult to do in the non-abelian case and to compare results in $SIM(2)$ formalism with the results in the usual Poincare-invariant formalism.

The super-Poincare abelian gauge theory contains a real scalar superfield $V$ and is invariant under the gauge transformation
\begin{equation}
\label{eq:abel:poincare_gauge_transformation}
 V' = V + i\al{\Lambda} -i\cl{\Lambda},
\end{equation}
where $\cl{\Lambda}$ is a chiral superfield.
In order to rewrite it in the $SIM(2)$ superspace formalism we define the $SIM(2)$ projections $\pptn{v}$, $\pptu{v}$, $\pptd{v}$, $\pptx{v}$, $\pptcn{\lambda}$, $\pptcu{\lambda}$, $\pptan{\lambda}$, $\pptad{\lambda}$ of $V$, $\cl{\Lambda}$, $\al{\Lambda}$ in the same way as we did in \eqref{eq:sim2:projections_redefined}. As a consequence of the reality of the superfield $V$ we have \eqref{eq:sim2:real_scalar_conjugation}, the chirality and antichirality of $\cl{\Lambda}$ and $\al{\Lambda}$ results in conditions \eqref{eq:sim2:chiral_conditions}, \eqref{eq:sim2:chiral_conjugation} with $\tilde{\lambda}$ in place of $\tilde{\phi}$.

The gauge transformation \eqref{eq:abel:poincare_gauge_transformation} rewritten for the $SIM(2)$ superfields is
\begin{align}
\label{eq:abel:gauge_transformation}
 \pptn{v}' &= \pptn{v} + i\pptan{\lambda} -i\pptcn{\lambda}, & 
 \pptu{v}' &= \pptu{v} - i\pptcu{\lambda}, \nonumber\\
 \pptx{v}' &= \pptx{v} + \frac{i}{2}\frac{\dbox}{i\dpp}\left(  \pptan{\lambda} + \pptcn{\lambda}  \right), & 
 \pptd{v}' &= \pptd{v} + i\pptad{\lambda}.
\end{align}

The action for the gauge field is
\begin{equation}
 S = \int\di^4x\di^2\cl{\theta} \, W^2
\end{equation}
where 
\begin{equation}
 W_{\ci{\alpha}} = i\al{\text{D}}^2\cl{\text{D}}_{\ci{\alpha}}V.
\end{equation}
In the $SIM(2)$ superspace formalism it looks like
\begin{equation}
\label{eq:abel:action_general}
 S = \int\di^4x\cd\ad\left[
 -\cd\left(\pptx{v} + \frac{1}{2}\frac{\dbox}{i\dpp}\pptn{v}\right) \ad\left(\pptx{v} - \frac{1}{2}\frac{\dbox}{i\dpp}\pptn{v}\right)
 + \cd\pptd{v}\frac{\dbox}{i\dpp}\ad\pptu{v}
 \right].
\end{equation}

It is possible to fix the gauge freedom in a way that respect $SIM(2)$ supersymmetry. A convenient choice of gauge fixing conditions, which completely fix the gauge freedom, is
\begin{align}
\label{eq:abel:gauge_fixing_conditions}
 \pptn{v} &= 0, & \cd\pptu{v} &= 0 = \ad\pptd{v}.
\end{align}
If $\pptn{v}$, $\pptu{v}$, $\pptd{v}$, $\pptx{v}$ are arbitrary then we can go to the above gauge by performing the gauge transformation \eqref{eq:abel:gauge_transformation} with
\begin{align}
 \pptcn{\lambda} &= -i\frac{\ad\cd}{i\dpp}\pptn{v}, &
 \pptcu{\lambda} &= -i\frac{\ad\cd}{i\dpp}\pptu{v}.
\end{align}
The only gauge transformations preserving our gauge fixing conditions are those with $\pptcn{\lambda} = 0 = \pptcu{\lambda}$, so there is no remaining gauge freedom.

In this gauge the action \eqref{eq:abel:action_general} reduces to
\begin{equation}
S = \int\di^4x\cd\ad\left(  -\cd\pptx{v}\ad\pptx{v} + \pptd{v}\dbox\pptu{v}  \right).
\end{equation}
In the gauge fixed form, the whole dynamics of super-Poincare abelian gauge theory is described by one real Grassmann even $SIM(2)$ superfield $\pptx{v}$ and one chiral Grassmann odd $SIM(2)$ superfield $\pptd{v}$ (and its conjugate $\pptu{v}$). 

The classical equations of motion are
\begin{align}
 \frac{\delta S}{\delta\pptx{v}} &= \cd\ad\pptx{v} - \ad\cd\pptx{v} = 0, & \frac{\delta S}{\delta\pptu{v}} &= \dbox\cd\pptd{v} = 0.
\end{align}
They are equivalent to
\begin{align}
\label{eq:abel:eom_susy}
 \dbox\pptd{v} &= 0, & \pptx{v} &= 0.
\end{align}
We see that $\pptx{v}$ is auxiliary superfield and all dynamics is carried by the superfield $\pptd{v}$ (which contains two bosonic and two fermionic degrees of freedom).

We will look how the above gauge and equations of motion look like if we rewrite them in terms of components of $V$
\begin{align}
 C &= V\vert, & 
  \cl{\chi}_\ci{\alpha} &= i\cl{D}_\ci{\alpha}V\vert, & 
  \al{\chi}_\ai{\alpha} &= -i\al{D}_\ai{\alpha}V\vert, \nonumber\\
 \cl{M} &= \cl{D}^2V\vert, & 
  \al{M} &= \al{D}^2V\vert, & 
  \vdd{A}{\alpha}{\alpha} &= \frac{1}{2}[\al{D}_\ai{\alpha},\cl{D}_\ci{\alpha}]V\vert, \nonumber\\
 \cl{\lambda}_\ci{\alpha} &= i\al{D}^2\cl{D}_\ci{\alpha}V\vert, & 
  \al{\lambda}_\ai{\alpha} &= -i\cl{D}^2\al{D}_\ai{\alpha}V\vert, &
  D' &= \frac{1}{2}\cl{D}^\ci{\alpha}\al{D}^2\cl{D}_\ci{\alpha}V\vert.
\end{align}
where $\vert=\vert_{\cl{\theta}^\ci{\alpha}=0,\al{\theta}^\ai{\alpha}=0}$ denotes projection, which leaves only $\theta$-independent part.

The gauge fixing conditions \eqref{eq:abel:gauge_fixing_conditions} imply
\begin{align}
 C &= 0, & \cl{\chi}_\ci{+} &= 0 = \al{\chi}_\ai{+}, & \vdd{A}{+}{+} &= 0, \nonumber\\
 \cl{M} &= 0 = \al{M}, & \cl{\lambda}_\ci{+} &= \dpp\al{\chi}_\ai{-}, & \al{\lambda}_\ai{+} &= \dpp\cl{\chi}_\ci{-}.
\end{align}
The last two conditions can be also written as $\cl{\chi}_\ci{-}=\frac{1}{\dpp}\al{\lambda}_\ai{+}$, $\al{\chi}_\ai{-}=\frac{1}{\dpp}\cl{\lambda}_\ci{+}$. We can write these conditions also in the form which does not depend on the choice of the vector $n$ as
\begin{align}
 C &= 0, & 
  \cl{M} &= 0 = \al{M}, & 
  n\cdot A &= 0, & 
  \chi &= i\frac{\fsn{n}}{2n\cdot\partial}\lambda, &
\end{align}
The only fields which are not completely constrained by these conditions are the scalar $D'$, the spinor $\cl{\lambda}$, and the vector $A$ which is constrained by $n\cdot A = 0$.
The components of the $SIM(2)$ superfields appearing in the action can be in the given gauge expressed as
\begin{align}
 \pptu{v}\vert &= \frac{1}{i\dpp}\al{\lambda}_\ai{+} = \frac{1}{i\dpp}\tilde{\al{\lambda}}_\ai{+}, &
 \pptd{v}\vert &= -\frac{1}{i\dpp}\cl{\lambda}_\ci{+} = -\frac{1}{i\dpp}\tilde{\cl{\lambda}}_\ci{+}, \nonumber\\
 \ad\pptu{v}\vert &= \vdd{A}{-}{+} = \vdd{\tilde{A}}{-}{+}, &
 \cd\pptd{v}\vert &= -\vdd{A}{+}{-} = -\vdd{\tilde{A}}{+}{-}, \nonumber\\
 \cd\pptx{v}\vert &= -\al{\lambda}_\ai{-} + \frac{\dpm}{\dpp}\al{\lambda}_\ai{+} = -\tilde{\al{\lambda}}_\ai{-}, &
 \ad\pptx{v}\vert &= -\cl{\lambda}_\ci{-} + \frac{\dmp}{\dpp}\cl{\lambda}_\ci{+} = -\tilde{\cl{\lambda}}_\ci{-}, \nonumber\\
 \frac{1}{2}[\cd,\ad]&\pptx{v}\vert = D', &
 \pptx{v}\vert = -\vdd{A}{-}{-} &+ \frac{\dmp}{\dpp}\vdd{A}{+}{-} + \frac{\dpm}{\dpp}\vdd{A}{-}{+} = -\vdd{\tilde{A}}{-}{-},
\end{align}
The equations of motion \eqref{eq:abel:eom_susy} written for the components are
\begin{align}
 \dbox\tilde{\cl{\lambda}}_\ci{+} &= 0 = \dbox\tilde{\al{\lambda}}_\ai{+}, & 
 \dbox\vdd{\tilde{A}}{+}{-} &= 0 = \dbox\vdd{\tilde{A}}{-}{+}, \nonumber\\
 \tilde{\cl{\lambda}}_\ci{-} &= 0 = \tilde{\al{\lambda}}_\ai{-}, &
 \vdd{\tilde{A}}{-}{-} &= 0, &
 D' &= 0.
\end{align}
Here we can identify the equation of motion of the massless fermion \eqref{eq:sim2:eom_ferminon_redefined} and the equation of motion of the abelian gauge field \eqref{eq:sim2:eom_maxwell}.


\section{Gauge chiral representation}
\label{sec:chiym}

This section is devoted to the non-abelian theory in the gauge chiral representation. We will show that it is possible to use the same $SIM(2)$ invariant gauge fixing conditions as in the case of the abelian theory to completely remove the gauge freedom. Then we will show how the action looks like in this gauge. We will not present the results for the case where the gauge is not fixed.

In the supersymmetric non-abelian gauge theory we have a chiral superfield $\cl{\Phi}$ with values in the representation space, and we require that it is invariant with respect to a local gauge transformation
\begin{equation}
 \cl{\Phi}' = e^{i\cl{\Lambda}}\cl{\Phi},
\end{equation}
where $\cl{\Lambda}$ is a Lie algebra valued chiral superfield, i.e. $\cl{\Lambda}=\cl{\Lambda}^AT_A$, where $\cl{\Lambda}^A$ are chiral superfields and $T_A$ are hermitian generators of Lie algebra. In order to construct a gauge invariant equivalent of the term $\al{\Phi}\cl{\Phi}$, we introduce a Lie algebra valued real scalar superfield $V$, which transforms as
\begin{equation}
\label{eq:chiral:poincare_gauge_transformation}
 e^{V'} = e^{i\al{\Lambda}}e^Ve^{-i\cl{\Lambda}},
\end{equation}
so the term $\al{\Phi}e^V\cl{\Phi}$ is gauge invariant.

The decomposition of the superfields $V$ and $\cl{\Lambda}$ to $SIM(2)$ superfields can be done in the same way as in the case of abelian gauge theory, i.e. according to \eqref{eq:sim2:projections_redefined}. The $SIM(2)$ superfields $\pptn{v}$, $\pptu{v}$, $\pptd{v}$, $\pptx{v}$ satisfy the conditions \eqref{eq:sim2:real_scalar_conjugation}, the $SIM(2)$ superfields $\pptcn{\lambda}$, $\pptcu{\lambda}$, $\pptan{\lambda}$, $\pptad{\lambda}$ satisfy the conditions \eqref{eq:sim2:chiral_conditions}, \eqref{eq:sim2:chiral_conjugation}.

Although it is not possible to rewrite the gauge transformation \eqref{eq:chiral:poincare_gauge_transformation} for $SIM(2)$ superfields in a compact form (at least we do not know how to do it), it is possible to fix the gauge in the same way as in the case of abelian theory, i.e. by requiring
\begin{align}
\label{eq:chiral:gauge_fixing_conditions}
 \pptn{v} &= 0, & \cd\pptu{v} &= 0 = \ad\pptd{v}.
\end{align}

Before we show that this choice of gauge is possible, we prove that an exponential of any complex $SIM(2)$ superfield $f$ can be written as
\begin{equation}
\label{eq:chiral:exp_def}
 e^f = e^{\al{a}}e^{\cl{c}},
\end{equation}
where $\cl{c}$ is a chiral $SIM(2)$ superfield and $\al{a}$ is an antichiral $SIM(2)$ superfield. This is very similar to the decomposition \eqref{eq:sim2:decomposition} which allows us to write any $SIM(2)$ superfield as a sum of a chiral and an antichiral $SIM(2)$ superfields. In the abelian case the decomposition \eqref{eq:chiral:exp_def} can be inferred directly from \eqref{eq:sim2:decomposition}.

First we will assume that the decomposition \eqref{eq:chiral:exp_def} is possible and find out what $\cl{c}$ and $\al{a}$ should be. Then we will show that this assumption is correct. A simple calculation shows, that the $SIM(2)$ superfields $\cl{c}$ and $\al{a}$ have to satisfy the equations
\begin{align}
\label{eq:chiral:exp_eq}
 \ad\left(e^{-f}\cd e^{f}\right) &= e^{-\cl{c}}i\dpp e^{\cl{c}}, & \cd\left(\ad e^{f}e^{-f}\right) &= i\dpp e^{\al{a}}e^{-\al{a}}.
\end{align}
The solutions of these equations can be formally written with the help of the path-ordering operators $\mathcal{R}$ and $\mathcal{L}$, which order the arguments according to increasing and decreasing value of $\vuu{x}{+}{+}$. The argument having the largest value of $\vuu{x}{+}{+}$ is the rightmost in the case of operator $\mathcal{R}$ and the leftmost in the case of the operator $\mathcal{L}$. The solutions are
\begin{align}
 e^{\cl{c}} &= \mathcal{R}\exp\left( -i\int_{-\infty}^{\vuu{x}{+}{+}} \ad\left(e^{-f}\cd e^{f}\right) \di\vuu{t}{+}{+} \right), \nonumber \\
 e^{\al{a}} &= \mathcal{L}\exp\left( -i\int_{-\infty}^{\vuu{x}{+}{+}} \cd\left(\ad e^{f}e^{-f}\right) \di\vuu{t}{+}{+} \right).
\end{align}
Now we define a new $SIM(2)$ superfield $f'$ as
\begin{equation}
 e^{f'} = e^{-\al{a}} e^f e^{-\cl{c}},
\end{equation}
In order to prove that the decomposition \eqref{eq:chiral:exp_def} is really possible, we have to prove that $f'=0$. With the help of \eqref{eq:chiral:exp_eq} it can be shown that
\begin{align}
 \ad\left(e^{-f'}\cd e^{f'}\right) &= 0, & \cd\left(\ad e^{f'}e^{-f'}\right) &= 0.
\end{align}
These equations are equivalent to
\begin{align}
 \ad\cd e^{f'} &= \ad e^{f'} e^{-f'} \cd e^{f'}, & \cd\ad e^{f'} &= - \ad e^{f'} e^{-f'} \cd e^{f'}.
\end{align}
Their sum gives us the equation $0 = (\ad\cd + \cd\ad)e^{f'} = i\dpp e^{f'}$, which is equivalent to $\dpp f' = 0$. The only solution of the last equation is $f'=0$ and this completes our proof.

Now we are going to prove that the gauge \eqref{eq:chiral:gauge_fixing_conditions} is admissible. We will show that by performing two subsequent gauge transformations we can go from arbitrary $SIM(2)$ superfields $\pptn{v}$, $\pptu{v}$, $\pptd{v}$, $\pptx{v}$ to superfields satisfying our gauge fixing conditions.

Let make a projection on $SIM(2)$ superspace on both sides of \eqref{eq:chiral:poincare_gauge_transformation}
\begin{equation}
 \left.\left(e^{V'}\right)\right\rVert = \left.\left(e^{i\al{\Lambda}}e^Ve^{-i\cl{\Lambda}}\right)\right\rVert
 \quad\Rightarrow\quad
 e^{\pptn{v}'} = e^{i\pptan{\lambda}}e^{\pptn{v}}e^{-i\pptcn{\lambda}}.
\end{equation}
We want to set $\pptn{v}'=0$, which will be achieved if $e^{\pptn{v}}=e^{-i\pptan{\lambda}}e^{i\pptcn{\lambda}}$. The decomposition \eqref{eq:chiral:exp_def} tells us, that this happens when we choose
\begin{equation}
 \pptcn{\lambda} = -i\ln\left( \mathcal{R}\exp\left( -i\int_{-\infty}^{\vuu{x}{+}{+}} \ad\left(e^{-\pptn{v}}\cd e^{\pptn{v}}\right) \di\vuu{t}{+}{+} \right) \right).
\end{equation}

Now we may assume that $\pptn{v}=0$ and perform another gauge transformation to set $\cd\pptu{v}=0$. In order to preserve the condition $\pptn{v}=0$ we have to choose a gauge transformation with $\pptcn{\lambda}=0$. By acting with $\cl{\text{D}}_{\ci{-}}$ on both sides of \eqref{eq:chiral:poincare_gauge_transformation} and then making projection on the $SIM(2)$ superspace we obtain
\begin{equation}
 \left.\cl{\text{D}}_{\ci{-}}\left(e^{V'}\right)\right\rVert = \left.{\text{D}}_{\ci{-}}\left(e^{i\al{\Lambda}}e^Ve^{-i\cl{\Lambda}}\right)\right\rVert
 \quad\Rightarrow\quad
 \pptu{v}' = \pptu{v} -i\pptcu{\lambda}.
\end{equation}
If we choose
\begin{equation}
 \pptcu{\lambda} = -i\frac{\ad\cd}{i\dpp}\pptu{v},
\end{equation}
we set $\cd\pptu{v}'=0$ (and also $\ad\pptd{v}'=0$).

This completes the proof of the admissibility of the gauge fixing conditions \eqref{eq:chiral:gauge_fixing_conditions}. Because the only gauge transformation which preserves our gauge fixing conditions is the one with $\pptcn{\lambda} = 0$, $\pptcu{\lambda} = 0$, the gauge freedom is fixed completely. 

The action for the non-abelian gauge field in the Poincare invariant formalism is
\begin{equation}
 S = \int\di^4x\di^2\cl{\theta} \, \text{tr}\left(W^2\right)
\end{equation}
where 
\begin{equation}
 W_{\ci{\alpha}} = i\al{\text{D}}^2\left(e^{-V}\cl{\text{D}}_{\ci{\alpha}}e^V\right).
\end{equation}

If we rewrite it in the $SIM(2)$ superspace formalism with the gauge being fixed according to \eqref{eq:chiral:gauge_fixing_conditions} then we get
\begin{equation}\begin{split}
\label{eq:chiral:gauge_sim_action}
 S = \int\di^4x\cd\ad \, \text{tr}\Bigg(
  & \pptd{v}\dbox\pptu{v} + \ad\left(\pptx{v}+\frac{1}{2}\{\pptu{v},\pptd{v}\}\right)\cd\left(\pptx{v}+\frac{1}{2}\{\pptu{v},\pptd{v}\}\right) \\
  & + \left(\pptx{v}+\frac{1}{2}\{\pptu{v},\pptd{v}\}\right)[\cd\pptd{v},\ad\pptu{v}] \\
  & - \cd\pptd{v}\left[\ad\pptu{v},\frac{\dpm}{\dpp}\ad\pptu{v}\right] - \ad\pptu{v}\left[\cd\pptd{v},\frac{\dmp}{\dpp}\cd\pptd{v}\right]
 \Bigg).
\end{split}\end{equation}
Note that each term in the sum is separately $SIM(2)$ invariant.

Now we will look what happens when the theory contains another field coupled to the gauge field. We have decided to use the model with a chiral field $\cl{\Phi}$ minimally coupled to the gauge field $V$, i.e. the action in the super-Poincare formalism is
\begin{equation}
\label{eq:chiral:poincare_matter_action}
 S = \int\di^4x\di^2\cl{\theta}\di^2\al{\theta} \, \al{\Phi}e^V\cl{\Phi}.
\end{equation}
The $SIM(2)$ superfields corresponding to the super-Poincare chiral field $\cl{\Phi}$ and its hermitian conjugate $\al{\Phi}$ are defined according to \eqref{eq:sim2:projections_redefined} and satisfies the conditions \eqref{eq:sim2:chiral_conditions}, \eqref{eq:sim2:chiral_conjugation}. When the action is rewritten in the $SIM(2)$ formalism it looks like
\begin{equation}\begin{split}
\label{eq:chiral:sim_matter_action}
 S = \int\di^4x\cd\ad \, \Bigg(
  & - \pptan{\phi}\frac{\dbox}{i\dpp}\pptcn{\phi} - \pptad{\phi}\pptcu{\phi} \\
  & - \pptad{\phi}\pptu{v}\pptcn{\phi} - \pptan{\phi}\pptd{v}\pptcu{\phi} + \pptan{\phi}\left(\pptx{v}+\frac{1}{2}(\pptu{v}\pptd{v}-\pptd{v}\pptu{v})\right)\pptcn{\phi} \\
  & + \left(\frac{\dpm}{\dpp}\pptan{\phi}\right)\left(\ad\pptu{v}\right)\pptcn{\phi} - \pptan{\phi}\left(\frac{\dpm}{\dpp}\ad\pptu{v}\right)\pptcn{\phi} \\
  & - \pptan{\phi}\left(\cd\pptd{v}\right)\left(\frac{\dmp}{\dpp}\pptcn{\phi}\right) + \pptan{\phi}\left(\frac{\dmp}{\dpp}\cd\pptd{v}\right)\pptcn{\phi}
 \Bigg).
\end{split}\end{equation}
Each term from the first and second row is separately $SIM(2)$ invariant. This is not true for terms from the last two rows where we have to group the terms in each row to get invariant expressions.

The formalism presented in this section have manifest $SIM(2)$ invariance but there is no gauge invariance because we are working in a gauge which completely removes it. This will be useful when we want to quantise it because there will not be any ghosts. But for other purposes it would be more beneficial if we had a formalism where the gauge freedom is not removed and where both $SIM(2)$ invariance and gauge invariance are manifest.

\subsection{Alternative decomposition of $V$ into $SIM(2)$ superfields}

We may also use the projections defined as
\begin{align}
\label{eq:wf:projections}
 &\begin{aligned}
 \ppn{\hat{v}} &= \left. V \right\rVert, \qquad &
 \ppu{\hat{v}} &= \left. e^{-V}\left(\cl{\text{D}}_{\ci{-}}e^V\right) \right \rVert, \qquad &
 \ppd{\hat{v}} &= \left. \left(\al{\text{D}}_{\ai{-}}e^V\right)e^{-V}\right \rVert,
 \end{aligned}\nonumber\\
 &\ppx{\hat{v}} = \left.\left( \frac{1}{2}\cl{\text{D}}_{\ci{-}}\left(\al{\text{D}}_{\ai{-}}e^Ve^{-V}\right)e^V
  - \frac{1}{2}e^V\al{\text{D}}_{\ai{-}}\left(e^{-V}\cl{\text{D}}_{\ci{-}}e^V\right) \right)\right\rVert, \nonumber\\
 &\begin{aligned}
 \ppn{\hat{\cl{\lambda}}} &= \cl{\Lambda}\rVert, \qquad&
 \ppu{\hat{\cl{\lambda}}} &= \left.i\left(\cl{\text{D}}_{\ci{-}}e^{-i\cl{\Lambda}}\right)e^{i\cl{\Lambda}}\right\rVert, \qquad&
 \ppn{\hat{\al{\lambda}}} &= \al{\Lambda}\rVert, \qquad&
 \ppd{\hat{\al{\lambda}}} &= \left.-ie^{-i\al{\Lambda}}\left(\al{\text{D}}_{\ai{-}}e^{i\al{\Lambda}}\right)\right\rVert.
 \end{aligned}
\end{align}
They satisfy the reality conditions \eqref{eq:sim2:real_scalar_conjugation}, but the chirality conditions for $\ppn{\hat{\cl{\lambda}}}$, $\ppu{\hat{\cl{\lambda}}}$, $\ppn{\hat{\al{\lambda}}}$, $\ppd{\hat{\al{\lambda}}}$ and infinitesimal $SIM(2)$ rotations are more complicated and contain derivatives acting on $e^V$. The gauge transformations are
\begin{align}
 e^{\ppn{\hat{v}'}} &= e^{i\ppn{\hat{\al{\lambda}}}}e^{\ppn{\hat{v}}}e^{-i\ppn{\hat{\cl{\lambda}}}}, \nonumber\\
 \ppu{\hat{v}'} &= e^{i\ppn{\hat{\cl{\lambda}}}}\left( \ppu{\hat{v}} - i\ppu{\hat{\cl{\lambda}}} \right)e^{-i\ppn{\hat{\cl{\lambda}}}}, \nonumber\\
 \ppd{\hat{v}'} &= e^{i\ppn{\hat{\al{\lambda}}}}\left( \ppd{\hat{v}} + i\ppd{\hat{\al{\lambda}}} \right)e^{-i\ppn{\hat{\al{\lambda}}}}, \nonumber\\
 \ppx{\hat{v}'} &= e^{i\ppn{\hat{\al{\lambda}}}}\ppx{\hat{v}}e^{-i\ppn{\hat{\cl{\lambda}}}} 
  + \frac{i}{2}\dmm e^{i\ppn{\hat{\al{\lambda}}}}e^{\ppn{\hat{v}}}e^{-i\ppn{\hat{\cl{\lambda}}}}
  - \frac{i}{2}e^{i\ppn{\hat{\al{\lambda}}}}e^{\ppn{\hat{v}}}\dmm e^{-i\ppn{\hat{\cl{\lambda}}}},
\end{align}
If we used the covariant derivatives $\tilde{\cl{D}}_\ci{-}$, $\tilde{\al{D}}_\ai{-}$ \eqref{eq:sim2:D_redefined} instead of $\cl{D}_\ci{-}$, $\al{D}_\ai{-}$ in the definition of projections \eqref{eq:wf:projections}, then the gauge transformations would be more complicated, but $SIM(2)$ projections would have nicer properties. The projections of $V$ would still satisfy the reality conditions \eqref{eq:sim2:real_scalar_conjugation}, the projections of $\cl{\Phi}$, $\al{\Phi}$ would satisfy chirality conditions \eqref{eq:sim2:chiral_conditions} and reality conditions \eqref{eq:sim2:real_scalar_conjugation}. The infinitesimal $SIM(2)$ rotations would look like \eqref{eq:sim2:rotations_redefined}.

\section{Elimination of the auxiliary superfields}
\label{sec:chiel}

How many real components, i.e. real fields with values in the representation space, are contained in the superfields $V$ and $\cl{\Phi}$? How many of them are physical, i.e. how many of them carry dynamics? There are 16 real components in $V$ and 8 real components in $\Phi$. Each $V$ and $\cl{\Phi}$ have four physical components (two bosonic and two fermionic).

The super-Poincare superfields were replaced by $SIM(2)$ superfields. The superfield $\cl{\Phi}$ was replaced by two $SIM(2)$ chiral superfields $\pptcn{\phi}$, $\pptcu{\phi}$, each having four real components. In the case of the superfield $V$ we have completely fixed the gauge by \eqref{eq:chiral:gauge_fixing_conditions} and only one real $SIM(2)$ superfield $\pptx{v}$ and one $SIM(2)$ chiral superfield $\pptu{v}$ (and its conjugate $\pptd{v}$) remains in the rewritten action. Each of $\pptx{v}$, $\pptu{v}$ has 4 real components so we got rid of 8 components of $V$.

In this section we will show that the $SIM(2)$ superfields $\pptcu{\phi}$ and $\pptx{v}$ are auxiliary and can be eliminated from the action. After that we will obtain an action containing only superfields whose components are physical.

We start by eliminating the auxiliary $SIM(2)$ superfield $\pptcu{\phi}$. By varying the action \eqref{eq:chiral:sim_matter_action} with respect to $\pptad{\phi}$ we obtain the equation of motion $\cd\pptcu{\phi} + \cd(\pptu{v}\pptcn{\phi}) = 0$. Its solution $\pptcu{\phi}=-\frac{1}{i\dpp}\ad\cd(\pptu{v}\pptcn{\phi})$ (and its conjugate) can be used to eliminate $\pptcu{\phi}$ and $\pptad{\phi}$ from the action. The resulting action is
\begin{equation}\begin{split}
\label{eq:auxiliary:sim_matter_action}
 S_c &= \int\di^4x\cd\ad \, \Bigg(
   - \pptan{\phi}\frac{\dbox}{i\dpp}\pptcn{\phi} + \pptan{\phi}\left(\pptx{v} + \frac{1}{2}\{\pptu{v},\pptd{v}\}\right)\pptcn{\phi}
   - \pptan{\phi}\left(\cd\pptd{v}\right)\frac{1}{i\dpp}\left(\left(\ad\pptu{v}\right)\pptcn{\phi}\right) \\
  & + \left(\frac{\dpm}{\dpp}\pptan{\phi}\right)\left(\ad\pptu{v}\right)\pptcn{\phi} - \pptan{\phi}\left(\frac{\dpm}{\dpp}\ad\pptu{v}\right)\pptcn{\phi} 
   - \pptan{\phi}\left(\cd\pptd{v}\right)\left(\frac{\dmp}{\dpp}\pptcn{\phi}\right) + \pptan{\phi}\left(\frac{\dmp}{\dpp}\cd\pptd{v}\right)\pptcn{\phi}
 \Bigg).
\end{split}\end{equation}

We generalise our model before we eliminate the auxiliary $SIM(2)$ superfield $\pptx{v}$ from it. Instead of a model with one chiral superfield $\cl{\Phi}$ we will consider a model with multiple chiral superfields $\cl{\Phi}^{(k)}$ coupled to the gauge superfield. The action will be
\begin{equation}
 S = S_g + \sum_{k}S_c^{(k)},
\end{equation}
where $S_g$ is the action \eqref{eq:chiral:gauge_sim_action} for the gauge superfield and $S_c^{(k)}$ are the actions \eqref{eq:auxiliary:sim_matter_action} with the $SIM(2)$ superfield $\pptcn{\phi}$ replaced by $\pptcn{\phi}^{(k)}$.
When the auxiliary $SIM(2)$ superfield $\pptx{v}$ is eliminated from this action we obtain the action 
\begin{equation}
 \mathcal{S} = \mathcal{S}_g + \sum_{k}\mathcal{S}_c^{(k)} + \sum_{k,l}\mathcal{S}_e^{(k,l)},
\end{equation}
where the part of the action containing only the gauge superfield is
\begin{equation}\begin{split}
\label{eq:auxiliary:action_gauge}
 \mathcal{S}_g = \int\di^4x\cd\ad \, \text{tr}\Bigg(
  & \pptd{v}\dbox\pptu{v} - \cd\pptd{v}\left[\ad\pptu{v},\frac{\dpm}{\dpp}\ad\pptu{v}\right] - \ad\pptu{v}\left[\cd\pptd{v},\frac{\dmp}{\dpp}\cd\pptd{v}\right] \\
  & - \left(\frac{\cd}{i\dpp}[\cd\pptd{v},\ad\pptu{v}]\right)\left(\frac{\ad}{i\dpp}[\cd\pptd{v},\ad\pptu{v}]\right)
 \Bigg),
\end{split}\end{equation}
the part containing chiral superfield coupled to the gauge superfield is
\begin{equation}\begin{split}
\label{eq:auxiliary:action_matter}
 \mathcal{S}_c^{(k)} = \int\di^4x\cd\ad \, \Bigg(
  & - \pptan{\phi}^{(k)}\frac{\dbox}{i\dpp}\pptcn{\phi}^{(k)} - \pptan{\phi}^{(k)}\left(\cd\pptd{v}\right)\frac{1}{i\dpp}\left(\left(\ad\pptu{v}\right)\pptcn{\phi}^{(k)}\right) \\
  & + \left(\frac{\dpm}{\dpp}\pptan{\phi}^{(k)}\right)\left(\ad\pptu{v}\right)\pptcn{\phi}^{(k)} - \pptan{\phi}^{(k)}\left(\frac{\dpm}{\dpp}\ad\pptu{v}\right)\pptcn{\phi}^{(k)} \\
  & - \pptan{\phi}^{(k)}\left(\cd\pptd{v}\right)\left(\frac{\dmp}{\dpp}\pptcn{\phi}^{(k)}\right) + \pptan{\phi}^{(k)}\left(\frac{\dmp}{\dpp}\cd\pptd{v}\right)\pptcn{\phi}^{(k)} \\
  & - \pptan{\phi}^{(k)}\left( \frac{\cd\ad-\ad\cd}{(i\dpp)^2}[\cd\pptd{v}\ad\pptu{v}] \right)\pptcn{\phi}^{(k)}
 \Bigg),
\end{split}\end{equation}
and there is also a part where the chiral superfields are mixed among each other
\begin{equation}
 \mathcal{S}_e^{(k,l)} = \int\di^4x\cd\ad \, \text{tr}\left(
  - \left(\frac{\cd}{i\dpp}\left(\pptcn{\phi}^{(k)}\pptan{\phi}^{(k)}\right)\right)\left(\frac{\ad}{i\dpp}\left(\pptcn{\phi}^{(l)}\pptan{\phi}^{(l)}\right)\right)
 \right).
\end{equation}

There are terms whose structure is different from the structure of the terms which were present in the original actions \eqref{eq:chiral:gauge_sim_action}, \eqref{eq:chiral:sim_matter_action}. In the original actions the nonlocal operator $\frac{1}{\dpp}$ always acted on terms composed of only one $SIM(2)$ superfield. This means that it is possible to have Feynman rules where each nonlocal operator is associated with only one leg of the vertex.
This is not true for the actions with eliminated auxiliary superfields. For example in the third term in \eqref{eq:auxiliary:sim_matter_action} the operator $\frac{1}{\dpp}$ acts on $\left(\ad\pptu{v}\right)\pptcn{\phi}$. In Feynman diagram with a vertex corresponding to such term, the nonlocal operator will not be associated with single but with a pair of legs.


\section{Covariant representation}
\label{sec:cov}

In this section we will develop another formulation of $SIM(2)$ gauge theory. It will be based on the covariant representation of the supersymmetric gauge theory. We will see that it is possible to develop a formalism in which both the $SIM(2)$ invariance and the gauge invariance are manifest. At the end of the section we will show how this formalism is related to the formalism developed in the previous section.

We begin with a brief presentation of the covariant representation of the super-Poincare gauge theory. In the covariant representation we use a covariantly chiral superfield $\cl{\bPhi}$, which satisfy the condition $\cadi{\alpha}\cl{\bPhi} = 0$ ($\ccdi{\alpha}\al{\bPhi} = 0$), instead of the chiral superfield $\cl{\Phi}$ used in the previous section. We will distinguish superfields used in the covariant representation from the superfields used in the previous section by typesetting them in bold letters. 

The covariant derivatives $\ccdi{\alpha}$, $\cadi{\alpha}$, $\cdd{\alpha}{\alpha}$ are subject to the commutation relations 
\begin{align}
\label{eq:covariant:poincare_commutators}
 \{\ccdi{\alpha},\ccdi{\beta}\} &= 0 = \{\cadi{\alpha},\cadi{\beta}\}, & 
 \{\ccdi{\alpha},\cadi{\alpha}\} &= i\cdd{\alpha}{\alpha}, \nonumber\\
 [\ccdi{\alpha},\cdd{\beta}{\beta}] &= C_{\ci{\alpha}\ci{\beta}}\al{\clW}_\ai{\beta}, & 
 [\cadi{\alpha},\cdd{\beta}{\beta}] &= C_{\ai{\alpha}\ai{\beta}}\cl{\clW}_\ci{\beta}, \nonumber\\
 [\cdd{\alpha}{\alpha},\cdd{\beta}{\beta}] &= -i(C_{\ai{\alpha}\ai{\beta}}\cl{\clF}_{\ci{\alpha}\ci{\beta}} + C_{\ci{\alpha}\ci{\beta}}\al{\clF}_{\ai{\alpha}\ai{\beta}}),
\end{align}
the field strengths $\cl{\clW}_\ci{\alpha}$, $\al{\clW}_\ai{\alpha}$, $\cl{\clF}_{\ci{\alpha}\ci{\beta}}$, $\al{\clF}_{\ai{\alpha}\ai{\beta}}$, $\clD'$ satisfy the relations
\begin{align}
\label{eq:covariant:poincare_identities}
 \cadi{\alpha}\cl{\clW}_\ci{\beta} &= 0, & 
 \ccdi{\alpha}\al{\clW}_\ai{\beta} &= 0, \nonumber\\
 \ccdi{\alpha}\cl{\clW}_\ci{\beta} &= \cl{\clF}_{\ci{\alpha}\ci{\beta}} - iC_{\ci{\alpha}\ci{\beta}}\clD', & 
 \cadi{\alpha}\al{\clW}_\ai{\beta} &= \al{\clF}_{\ai{\alpha}\ai{\beta}} + iC_{\ai{\alpha}\ai{\beta}}\clD'.
\end{align}

The superfield $\cl{\bPhi}$ and the covariant derivatives transform under the gauge transformation as
\begin{align}
 \cl{\bPhi}' &= e^{i\boldsymbol{K}}\cl{\bPhi}, &
 \ccdi{\alpha}' &= e^{i\boldsymbol{K}} \ccdi{\alpha} e^{-i\boldsymbol{K}}, &
 \cadi{\alpha}' &= e^{i\boldsymbol{K}} \cadi{\alpha} e^{-i\boldsymbol{K}}, &
 \cdd{\alpha}{\alpha}' &= e^{i\boldsymbol{K}} \cdd{\alpha}{\alpha} e^{-i\boldsymbol{K}},
\end{align}
where $\boldsymbol{K}$ is a real Lie algebra valued superfield.

The $SIM(2)$ projections of the superfields $\cl{\bPhi}$, $\al{\bPhi}$ are defined in a similar way as in the case of the superfields $\cl{\Phi}$, $\al{\Phi}$, the main difference is that we use the covariant derivatives instead of the ordinary ones. The covariant $SIM(2)$ projections
\begin{align}
\label{eq:covariant:phi_projections}
 \pptcn{\bphi} &= \cl{\bPhi}\rVert, &
 \pptcu{\bphi} &= \left.\left(\ccdi{-} - \cdd{-}{+}\frac{1}{\cdd{+}{+}}\ccdi{+}\right)\cl{\bPhi}\right\rVert, \nonumber\\
 \pptan{\bphi} &= \al{\bPhi}\rVert, &
 \pptad{\bphi} &= \left.\left(\cadi{-} - \cdd{+}{-}\frac{1}{\cdd{+}{+}}\cadi{+}\right)\al{\bPhi}\right\rVert,
\end{align}
satisfy the covariant $SIM(2)$ chirality conditions
\begin{align}
 \cad\pptcn{\bphi} &= 0 = \ccd\pptan{\bphi}, & \cad\pptcu{\bphi} &= 0 = \ccd\pptad{\bphi}.
\end{align}
Note that the ordering of the covariant derivatives in \eqref{eq:covariant:phi_projections} is important because not all of them commute among themselves. If the covariant derivatives $\cdd{-}{+}$, $\cdd{+}{-}$ were not placed in front of the other derivatives then the $SIM(2)$ projections $\pptcu{\bphi}$, $\pptad{\bphi}$ would not satisfy the covariant $SIM(2)$ chirality conditions. 
The projections $\pptcn{\bphi}$, $\pptcu{\bphi}$, $\pptan{\bphi}$, $\pptad{\bphi}$, to which we will refer also as to $SIM(2)$ superfields, transforms with respect to the infinitesimal $SIM(2)$ transformations exactly as the field $f$ in \eqref{eq:sim2:rotations_redefined}. The gauge transformations acts on them as 
\begin{align}
 \pptcn{\bphi}' &= e^{i\boldsymbol{k}}\pptcn{\bphi}, &
 \pptcu{\bphi}' &= e^{i\boldsymbol{k}}\pptcu{\bphi},
\end{align}
where the real $SIM(2)$ superfield $\boldsymbol{k} = \boldsymbol{K}\rVert$ is the projection of the superfield $\boldsymbol{K}$
\footnote{
When we operate in the $SIM(2)$ superspace we should use the $SIM(2)$ projections $\ccd\rVert$, $\cad\rVert$, $\cdd{\alpha}{\alpha}\rVert$ of the covariant derivatives which do not contain unwanted Grassmann variables $\cl{\theta}^\ci{-}$, $\al{\theta}^\ai{-}$. We will not distinguish the $SIM(2)$ projections of covariant derivatives from their unprojected super-Poincare counterparts. It should be clear from the context which derivatives should be used, moreover this difference is not important in most cases. The $SIM(2)$ projected covariant derivatives transform under the gauge transformation as
\begin{align}
 \ccd' &= e^{i\boldsymbol{k}} \ccd e^{-i\boldsymbol{k}}, &
 \cad' &= e^{i\boldsymbol{k}} \cad e^{-i\boldsymbol{k}}, &
 \cdd{\alpha}{\alpha}' &= e^{i\boldsymbol{k}} \cdd{\alpha}{\alpha} e^{-i\boldsymbol{k}}.
\end{align}
}.

The actions for the scalar chiral field and for the gauge field
\begin{align}
\label{eq:covariant:poincare_action}
 S_{scalar} &= \int\di^4x\di^2\cl{\theta}\di^2\al{\theta} \, \al{\bPhi}\cl{\bPhi}, &
 S_{gauge} &= \int\di^4x\di^2\cl{\theta} \, \text{tr}\left(\clW^2\right),
\end{align}
rewritten in the $SIM(2)$ superspace formalism they become
\footnote{
The $SIM(2)$ superfields $\pptcu{\bphi}$ and $\pptad{\bphi}$ are auxiliary and can be eliminated from the action. If we do that we obtain the action
\begin{equation}
 S = \int\di^4x\ccd\cad \, \left(
   \pptan{\bphi}\frac{\cad}{\cdd{+}{+}}\left( \dbox_{cov} + \cld' - \cl{\clw}_\ci{+}\frac{1}{i\cdd{+}{+}}\al{\clw}_\ai{+} \right)\frac{\ccd}{\cdd{+}{+}}\pptcn{\bphi}
 \right).
\end{equation}
}
\begin{align}
\label{eq:covariant:matter_action}
 S_{scalar} &= \int\di^4x\ccd\cad \, \Bigg(
   - \pptad{\bphi}\pptcu{\bphi} - \pptan{\bphi}\frac{\cad}{\cdd{+}{+}}\left(  \dbox_{cov} + \cld'  \right)\frac{\ccd}{\cdd{+}{+}}\pptcn{\bphi} \\
  & \hspace{8em} + \pptad{\bphi}\al{\clw}_\ai{+}\frac{1}{i\cdd{+}{+}}\pptcn{\bphi} + \pptan{\bphi}\frac{1}{i\cdd{+}{+}}\left(\cl{\clw}_\ci{+}\pptcu{\bphi}\right)
 \Bigg), \nonumber\\
\label{eq:covariant:gauge_action}
 S_{gauge} &= \int\di^4x\ccd \text{tr}\Big(
  - \cl{\clw}_\ci{-}\left(\cadd{\clf}{+}{-} + \cld'\right) + \cl{\clw}_\ci{+}\cadd{\clf}{-}{-}
 \Big).
\end{align}
where $\dbox_{cov} = \frac{1}{2}\cdu{\alpha}{\alpha}\cdd{\alpha}{\alpha}$ is d'Alembertian composed of covariant derivatives and
\begin{align}
 \cl{\clw}_\ci{+} &= \cl{\clW}_\ci{+}\rVert, & 
 \al{\clw}_\ai{+} &= \al{\clW}_\ai{+}\rVert, & 
 \cld' &= \clD'\rVert, &
 \cadd{\clf}{+}{-} &= \cadd{\clF}{+}{-}\rVert, &
 \cadd{\clf}{-}{-} &= \cadd{\clF}{-}{-}\rVert,
\end{align}
are the $SIM(2)$ projections of the corresponding super-Poincare field strengths. It is easy to see that each term in \eqref{eq:covariant:matter_action} and \eqref{eq:covariant:gauge_action} is gauge invariant. The $SIM(2)$ invariance of \eqref{eq:covariant:matter_action} follows directly from the transformation rules for the superfields and the derivatives appearing in it. In order to prove the $SIM(2)$ invariance of \eqref{eq:covariant:gauge_action} we have to transform each subscript in the superfields $\cl{\clw}_\ci{+}$, $\cl{\clw}_\ci{-}$, $\cadd{\clf}{+}{-}$, $\cadd{\clf}{-}{-}$ according to \eqref{eq:sim2:sim2_rotations_matrix} and then use the identities $\ccd\cl{\clw}_\ci{-} = \cadd{\clf}{+}{-}-\cld'$, $\ccd\cl{\clw}_\ci{+}=\cadd{\clf}{+}{+}$.

Now we will describe how the covariant representation can be transformed to the gauge chiral representation described in the previous section. If we are working in the super-Poincare formalism and want to go from the covariant representation to the gauge chiral representation then we have to do the following replacements for the covariant derivatives and superfields
\begin{align}
 \ccdi{\alpha} &\rightarrow e^{-V}\cl{D}_\ci{\alpha}e^V, & 
 \cadi{\alpha} &\rightarrow \al{D}_\ai{\alpha}, & 
 \cdd{\alpha}{\alpha} &\rightarrow -i\{e^{-V}\cl{D}_\ci{\alpha}e^V,\al{D}_\ai{\alpha}\}, \nonumber\\
 \cl{\bPhi} &\rightarrow \cl{\Phi}, & 
 \al{\bPhi} &\rightarrow \al{\Phi}e^V.
\end{align}
In the $SIM(2)$ formalism we have to replace the covariantly chiral $SIM(2)$ superfields \eqref{eq:covariant:phi_projections} and covariant derivatives with expressions containing the chiral $SIM(2)$ superfields $\ppcn{\phi}$, $\ppcu{\phi}$, ordinary derivatives and the gauge fixed superfields $\pptu{v}$, $\pptd{v}$, $\pptx{v}$. 
\begin{align}
\label{eq:covariant:to_chiral}
&\begin{aligned}
 \ccd &\rightarrow \cd, & 
 \cad &\rightarrow \ad, \\
 \cdd{+}{+} &\rightarrow \dpp, &
 \qquad\qquad \cdd{-}{+} &\rightarrow \dmp - i\ad\pptu{v}, &
 \qquad\qquad \cdd{+}{-} &\rightarrow \dpm + i\cd\pptd{v},
\end{aligned} \nonumber\\
 &\cdd{-}{-} \rightarrow \dmm + i\pptx{v} - i\frac{\dpm}{\dpp}\ad\pptu{v} + i\frac{\dmp}{\dpp}\cd\pptd{v} + \frac{i}{2}\{\pptu{v},\pptd{v}\}, \nonumber\\
&\begin{aligned}
 \quad\; \pptcn{\bphi} &\rightarrow \pptcn{\phi}, & 
 \qquad\qquad\qquad \pptcu{\bphi} &\rightarrow \pptcu{\phi} + \pptu{v}\pptcn{\phi} - (\ad\pptu{v})\frac{\cd}{i\dpp}\pptcn{\phi}, \\
 \pptan{\bphi} &\rightarrow \pptan{\phi}, &
 \pptad{\bphi} &\rightarrow \pptad{\phi} + \pptan{\phi}\pptd{v} - \left(\frac{\ad}{i\dpp}\pptan{\phi}\right)(\cd\pptd{v}),
\end{aligned}
\end{align}
The field strengths appearing in \eqref{eq:covariant:matter_action}, \eqref{eq:covariant:gauge_action} can be calculated with the help of the commutation relations \eqref{eq:covariant:poincare_commutators} and the identities \eqref{eq:covariant:poincare_identities} as
\begin{align}
\label{eq:covariant:to_chiral_fields}
 \cl{\clw}_\ci{+} &= i[\cad,\cdd{+}{-}], & 
 \al{\clw}_\ai{+} &= i[\ccd,\cdd{-}{+}], \nonumber\\ 
 \cadd{\clf}{+}{-} &= \frac{1}{2}\left( [\cdd{-}{-},\cdd{+}{+}] + [\cdd{+}{-},\cdd{-}{+}] \right), &
 \cadd{\clf}{-}{-} &= [\cdd{-}{-},\cdd{-}{+}], \nonumber\\
 \cl{\clw}_\ci{-} &= i[\cad,\cdd{-}{-}], & 
 \al{\clw}_\ai{-} &= i[\ccd,\cdd{-}{-}], \nonumber\\
 \cld' &= \frac{1}{2}\left( \cad\al{\clw}_\ai{-} - \ccd\cl{\clw}_\ci{-} + [\cdd{+}{-},\cdd{-}{+}] \right).
\end{align}

All actions which have been presented so far respect super-Poincare symmetry even if they are written in the $SIM(2)$ formalism, which has lower symmetry. In order to break the Lorentz symmetry we have to add some Lorentz breaking terms. It seems that the covariant representation is most suitable for constructing such terms because it allows us to easily verify both the gauge and $SIM(2)$ invariance.


 
\section{Lorentz breaking terms}
\label{sec:break}

In this section a brief discussion of effects of $SIM(2)$ invariant Lorentz breaking mass terms on results presented in previous sections is given.

The following $SIM(2)$ invariant but Lorentz breaking mass term can be added to the action for the gauge field
\begin{equation}
 S_{mass-gauge} = m_g^2 \int\di^4x\ccd\cad\, \text{tr}\left( \al{\clw}_\ai{+} \frac{1}{(i\cdd{+}{+})^2} \cl{\clw}_\ci{+} \right).
\end{equation}
It can be rewritten in the gauge chiral representation if the substitution \eqref{eq:covariant:to_chiral}, \eqref{eq:covariant:to_chiral_fields} is performed as
\begin{equation}
 S_{mass-gauge} = - m_g^2 \int\di^4x\cd\ad\, \text{tr}\left( \pptd{v}\pptu{v} \right).
\end{equation}
Instead of the term $\pptd{v}\dbox\pptu{v}$ in \eqref{eq:chiral:gauge_sim_action}, and in \eqref{eq:auxiliary:action_gauge} we now have $\pptd{v}(\dbox-m_g^2)\pptu{v}$ so the $SIM(2)$ chiral superfield $\pptd{v}$ is now massive with mass $m_g$.
Note that the mass term does not break the gauge invariance and there are still two bosonic physical degrees of freedom in gauge multiplet. If we had added Lorentz invariant mass term for the gauge field, the gauge invariance would have been broken and there would have been three bosonic physical degrees of freedom.

The following $SIM(2)$ invariant but Lorentz breaking mass term can be added to the action for the chiral field
\begin{equation}
 S_{mass-chiral} = m_c^2 \int\di^4x\ccd\cad\, \left( \pptan{\bphi} \frac{1}{i\cdd{+}{+}} \pptcn{\bphi} \right).
\end{equation}
It can be rewritten in gauge chiral representation if the substitution \eqref{eq:covariant:to_chiral} is performed as
\begin{equation}
 S_{mass-chiral} = m_c^2 \int\di^4x\cd\ad\, \left( \pptan{\phi} \frac{1}{i\dpp} \pptcn{\phi} \right).
\end{equation}
The effect on \eqref{eq:chiral:sim_matter_action}, \eqref{eq:auxiliary:sim_matter_action} and \eqref{eq:auxiliary:action_matter} is that instead of the term $\pptan{\phi}\frac{\dbox}{i\dpp}\pptcn{\phi}$ we now have $\pptan{\phi}\frac{\dbox-m_c^2}{i\dpp}\pptcn{\phi}$ thus the $SIM(2)$ chiral superfield $\pptcn{\phi}$ has now mass $m_c$.

Both the mass term for gauge field and mass term for the chiral field were also proposed in \cite{uam}.


\section{Conclusions}

The model investigated in this paper consists of a gauge superfield minimally coupled to chiral superfields. 
Two formulations of the supersymmetric gauge theory in $SIM(2)$ superspace were presented. The first was based on the gauge chiral representation of the supersymmetric gauge theory, the second was based on the covariant representation of supersymmetric gauge theory.

A key observation which allowed us to reformulate the gauge chiral representation in $SIM(2)$ formalism is that it is possible to completely fix the gauge in a way that does not break $SIM(2)$ supersymmetry. While in the abelian case treated in section \ref{sec:chiab} we were able to obtain results without gauge fixing, in the non-abelian case treated in section \ref{sec:chiym} we were able to obtain results in a compact form only if the gauge was fixed.
While the gauge fixed theory does not have any gauge invariance it posses manifest $SIM(2)$ invariance and if we used it as a starting point for quantisation we would obtain a theory without ghosts.

The number of $SIM(2)$ superfields appearing in the theory has been reduced when we removed gauge freedom. Moreover in section \ref{sec:chiel} it was shown that some of the remaining $SIM(2)$ superfields are auxiliary and can be eliminated from the theory. This further reduces the number of $SIM(2)$ superfields appearing in the theory but the resulting action contains terms which would lead to more complicated Feynman rules.

While the gauge fixed gauge chiral representation gives us a theory suitable for quantisation, the covariant representation presented in section \ref{sec:cov} can be rewritten in $SIM(2)$ formalism in such a way that both $SIM(2)$ invariance and gauge invariance are manifest. Thus it is more suitable for theoretical considerations.
It was described how to rewrite the expressions from the covariant representation to the gauge fixed gauge chiral representation in the $SIM(2)$ formalism.

The Lorentz breaking but $SIM(2)$ invariant mass terms can be added to the gauge superfield or to the chiral superfield. Their effects on results of previous sections was discussed in section \ref{sec:break}.

\acknowledgments

The discussions with Rikard von Unge and Stanislav Petr\'a\v{s} are gratefully acknowledged. The research was supported by the Czech government grant agency under contract no. GACR 202/08/H072.



\end{document}